%% file: ag_dbexp_fpcp2014.tex
\documentclass[12pt]{article}
\usepackage{graphicx}
\usepackage{lineno}

\usepackage{hyperref}

\newcommand{\hepph}[1]{\href{http://www.arxiv.org/abs/hep-ph/#1}{\tt hep-ph/#1}}
\newcommand{\arXiv}[1]{\href{http://www.arxiv.org/abs/arXiv:#1}{\tt arXiv/#1}}
\newcommand{\hepex}[1]{\href{http://www.arxiv.org/abs/hep-ex/#1}{\tt hep-ex/#1}}
\newcommand{\nuclex}[1]{\href{http://www.arxiv.org/abs/nucl-ex/#1}{\tt nucl-ex/#1}}

\newcommand\pubnumber{}
\newcommand\pubdate{\today}

\def\unipd{Dipartimento di Fisica e Astronomia\\
via Marzolo, 11, I-35131 Padova, Italy\\}
\def\infnpd{Istituto Nazionale di Fisica Nucleare, Sezione di Padova\\
via Marzolo, 11, I-35131 Padova, Italy\\}
\def\email{\footnote{e-mail: alberto.garfagnini@pd.infn.it}}

\def\Title#1{\begin{center} {\Large #1 } \end{center}}
\def\Author#1{\begin{center}{ \sc #1} \end{center}}
\def\Address#1{\begin{center}{ \it #1} \end{center}}

\newcommand\pubblock{\rightline{\begin{tabular}{l} \pubnumber\\
         \pubdate  \end{tabular}}}
\newenvironment{Abstract}{\begin{quotation}  }{\end{quotation}}
\newenvironment{Presented}{\begin{quotation} \begin{center} 
             PRESENTED AT\end{center}\bigskip 
      \begin{center}\begin{large}}{\end{large}\end{center} \end{quotation}}

\input econfmacros.tex

\def\G          {{\mbox{{\sc Gerda}}}}
\def\expos      {kg$\cdot$yr}

\def\ctsper     {cts/(keV$\cdot$kg$\cdot$yr)}

\begin{document}

\begin{titlepage}
\pubblock

\vfill
\Title{Neutrinoless Double Beta Decay Experiments}
\vfill
\Author{Alberto Garfagnini\email}
\Address{\unipd\infnpd}
\vfill
\begin{Abstract}
Neutrinoless double beta decay is the only process known so far able to
test the neutrino intrinsic nature: its experimental observation
would imply that the lepton number is violated by two units and prove that
neutrinos have a Majorana mass components, being their own anti-particle.
While several experiments searching for such a rare decay
have been performed in the past, a new generation
of experiments using different isotopes and techniques have recently
released their results or are taking data and will provide new limits,
should no signal be observed, in the next few years to come.
The present contribution reviews the latest public results
on double beta decay searches and gives an overview on the
expected sensitivities
of the experiments in construction which will be able to set stronger limits
in the near future.
\end{Abstract}
\vfill
\begin{Presented}
2014 Flavor Physics and CP Violation (FPCP-2014), \\Marseille
France, May 26- 30 2014
\end{Presented}
\vfill
\end{titlepage}
\def\thefootnote{\fnsymbol{footnote}}
\setcounter{footnote}{0}

\section{Introduction}
Double beta decay is the simultaneous beta decay of two neutrons in a nucleus.
The reaction can be calculated in the standard model of particle physics as
a second order process:
$(A,Z) \rightarrow (A,Z+2) + 2 e^{-} + 2 \overline{\nu}_{e}$.
The two neutrino double beta decay ($2\nu\beta\beta$)
has been observed in eleven nuclei, where single beta decay is
energetically forbidden, and very high half-lives,
between $7 \times 10^{18}~\mbox{yr}$ and $2 \times 10^{24}~\mbox{yr}$
have been measured~\cite{bib:2nubb_pdg,bib:2nubb:rev_bara}.
Several models extending the standard model predict that
a neutrinoless double beta decay ($0\nu\beta\beta$) should also exists:
$(A,Z) \rightarrow (A,Z+2) + 2 e^{-}$. It's observation would imply that
lepton number is violated by two unit and that neutrinos have a Majorana
mass component. The standard mechanism for 
$0\nu\beta\beta$ assumes that
the process is mediated by light and massive Majorana neutrinos
and that other mechanisms potentially leading to neutrinoless double
beta decay are negligible~\cite{bib:0nubb:rode_standard_inter}.
With a light Majorana neutrino exchange,
it would be possible to derive an effective neutrino mass
using nuclear matrix element and phase space factor predictions.
For recent reviews on the subject,
see
Ref.~\cite{bib:2nubb:bernard_review},\cite{bib:2nubb:rodejohann_review}.\\
Several experiments searching for neutrinoless double beta decay have
been performed in the past, going back
for at least half a century, with increasing sensitivities.
Two main approaches have been followed. Indirect methods are based on
the measurements of anomalous concentrations of the daughter nuclei in
selected samples after very long exposures (i.e. radiochemical methods).
Direct methods, on the other hand, try to measure in real time
the properties of the two electrons emitted in $\beta\beta$ decay.
The detectors can be homogeneous, when the $\beta\beta$ source is the detector
medium and in-homogeneous when external $\beta\beta$ sources are inserted
in the detector.

In the following sections, the status of the art on $0\nu\beta\beta$
searches will be presented with special emphasis on large scale
running experiments.

%

%

\section{Double Beta Decay in $^{136}$Xe}
\label{sec:136xe}
$^{136}$Xe is a very interesting double beta decay emitter candidate.
It has a high $Q_{\beta\beta} = 2457~\mbox{keV}$, in a region which can have
lower contaminations from radioactive background events. It can can be
dissolved in liquid scintillators or used as gas allowing to realize
a homogeneous detector providing both scintillation and ionization signals.
Two large experiments have searched for $0\nu\beta\beta$ in Xe:
EXO-200 has used xenon in an homogeneous medium (both as $0\nu\beta\beta$
source and as detector), while in KamLAND-Zen it has been dissolved as a passive
$\beta\beta$ source in a liquid scintillator detector.

\subsection{EXO}
\label{sec:exo}
The Enriched Xenon Observatory~\cite{bib:EXO_jinst} is an experiment
in operation at the Waste Isolation Pilot Plant (WIPP), at a depth of
about 1600~m water equivalent near Carlsbad in New Mexico (USA).
The experiment is built around a large liquid Xenon Time Projection Chamber
filled with about 200~kg of liquid Xenon enriched to about 80.6\%
in the $^{136}$Xe isotope. In contrast to standard TPCs, the experiment uses
liquid xenon which can be concentrated in a smaller volume with the same
mass concentration. To overcome the limitation of worse energy resolutions
compared to gaseous TPCs, the experiment exploits the readout of both
scintillation and ionization signals produced by interacting particles
in xenon. Moreover, by combining both signals (scintillation light and
ionization charges), the experiment is able to
reject background events characterized by different charge to light collection
ratio. Finally, by using the difference in the arrival time between the
scintillation and ionization signals a z-coordinate of the event is
reconstructed.
The experiment started data taking in May 2011.

\begin{figure}[thb]
\begin{center}
\includegraphics[width=0.43\textwidth]{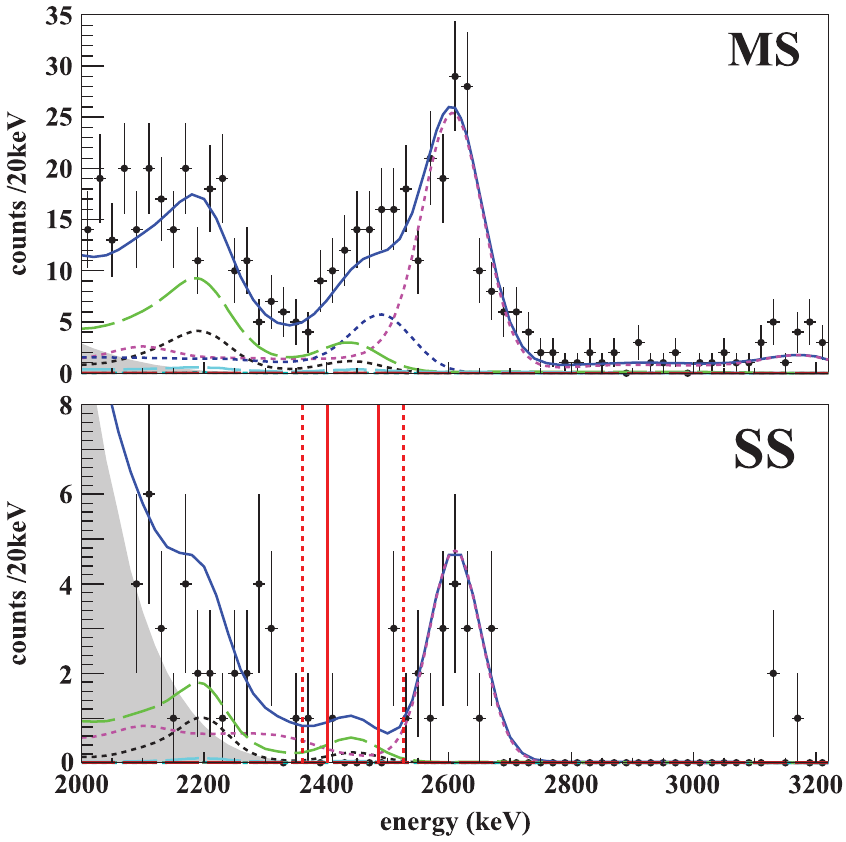}
\includegraphics[width=0.56\textwidth]{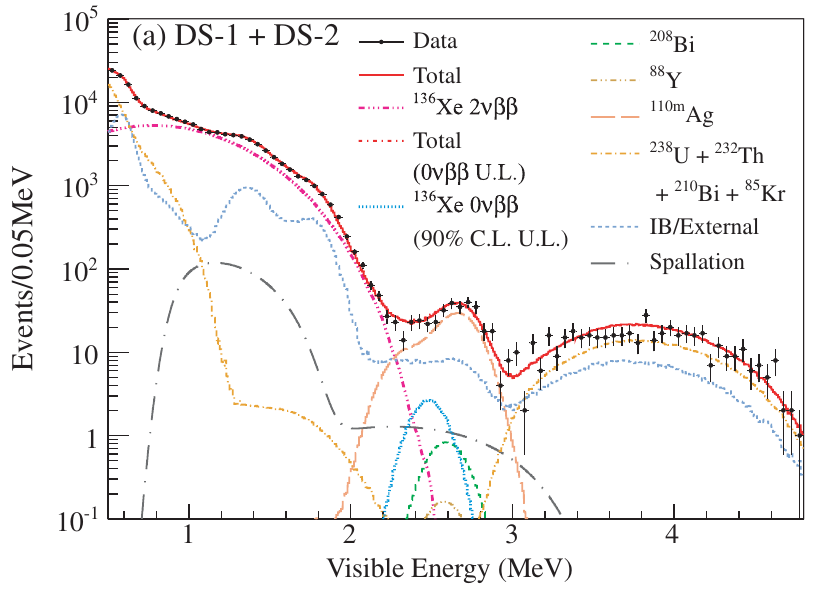}
\end{center}
\caption{Left: EXO-200. Energy spectra in the $^{136}$Xe $Q_{\beta\beta}$ region
for Multiple Site (top) and Single Site (bottom) events. The 1 (2) $\sigma$
region around $Q_{\beta\beta}$ is shown by solid (dashed) vertical
lines. Taken from Ref.~\protect\cite{bib:EXO_0nubb}.\quad
Right: KamLAND-Zen. Energy spectra of selected candidate events together
with the best-fit backgrounds and $2\nu\beta\beta$ decays. Figure taken
from Ref.~\protect\cite{bib:kamlandzen:0nubb}.}
\label{fig:exo:spectrum_0nubb}
\end{figure}

In June 2012, the collaboration reported the first results on $0\nu\beta\beta$
decay, analyzing and exposure of 32.5~kg$\cdot$yr. The $2\nu\beta\beta$ and
$0\nu\beta\beta$ signals were extracted simultaneously with a fit to the
single-site (SS) and multiple-site (MS) spectra, in an energy range
between 0.7~MeV and 3.5~MeV
(see left plots of Figure~\ref{fig:exo:spectrum_0nubb}).
The fit takes into consideration the main
radioactive background sources and a lower limit to the $0\nu\beta\beta$
life-time has been derived~\cite{bib:EXO_0nubb}:
$$T_{1/2}^{0\nu} > 1.6 \times 10^{25}~\mbox{yr}~~ @ ~90\% ~~\mbox{C.L. .}$$

The measurement has been recently updated with a higher exposure (100~kg~yr)
and with an improved detection sensitivity~\cite{bib:EXO_0nubb:nature}.
The claimed half-life sensitivity, $1.9\cdot10^{25}~\mbox{yr}$, is an
improvement by a factor of 2.7 compared to previous EXO-200 results.
They find no statistically significant evidence for $0\nu\beta\beta$ decay
but set a worse half-life limit of $1.1\cdot10^{25}$ yr at 90\% CL.

Very recently the EXO collaboration has published and update
on $2\nu\beta\beta$ life-time measurement~\cite{bib:EXO_2nubb_second}
using 127.6 days of live-time. The measurement,
$$T_{1/2}^{2\nu} = 2.165 \pm 0.016 \mbox{(stat.)}
                   \pm 0.059 \mbox{(syst.)} \times10^{21}~\mbox{yr}~\mbox{,}$$
is the most precisely measured half-life $2\nu\beta\beta$ decay to date.

A future evolution of EXO~\cite{bib:nEXO}
is moving in the direction of a tonne scale
experiment, with an active mass of few tonnes of $^{136}$Xe and improved
energy resolution and background suppression.

\subsection{KamLAND-Zen}
\label{sec:kamland-zen}
The KamLAND-Zen experiment is based on a
modification of the existing KamLAND~\cite{bib:kamland_jinst} detector
carried out in the summer of 2011. KamLAND is located at a depth of about
2700~m water equivalent at the Kamioka underground neutrino observatory
near Toyama in Japan.
The experiment has been equipped with
13 tons of Xe-loaded liquid scintillator (Xe-LS)
contained inside a 3.08~m diameter spherical inner balloon.
The isotopic abundance of the enriched xenon has been measured to be
about 90.9\% $^{136}$Xe and 8.9\% $^{134}$Xe.
The experiment started data taking in October 2011 and after an exposure of
30.8~kg~yr of $^{136}$Xe (77.6 days) it reported a measurement of
the $2\nu\beta\beta$ half-life\cite{bib:kamlandzen:2nubb}:
$$T_{1/2}^{2\nu} = 2.38 \pm 0.02 (\mbox{stat}) \pm 0.40 (\mbox{syst}).$$ 
Careful studies have been performed by the collaboration to identify
the various background sources contributing to the energy spectra (see
right plot of Figure~\ref{fig:exo:spectrum_0nubb}). The spectrum shows
a clear peak in the ROI that is compatible with a
$^{110m}$Ag contamination of the inner balloon. An attempt to purify the
Xe-LS has been performed, but unfortunately the filtration did not
produce the desired effect: the background counting rate due to 
$^{110m}$Ag decreased only slightly, from $0.19\pm 0.02$~cts/(tonne$\cdot$day)
to $0.14\pm 0.03$~cts/(tonne$\cdot$day).

The $0\nu\beta\beta$ limit reported so far by the
experiment is\cite{bib:kamlandzen:0nubb}:
$$T_{1/2}^{0\nu} > 1.9 \times 10^{25}~\mbox{yr}~~ @ ~90\% ~~\mbox{C.L.}$$

A careful and long lasting (1.5~yr) purification activity has allowed to
reduce radioactive impurities and to start a new data taking phase at the
end of 2013. A preliminary analysis of both running
phases~\cite{bib:kamlandzen:nu2014} improves the current $0\nu\beta\beta$
limit to
$$T_{1/2}^{0\nu} > 2.6 \times 10^{25}~\mbox{yr}~~ @ ~90\% ~~\mbox{C.L.}$$

Several detector improvements are foreseen in the years to come and an
increase in the Xe mass (up to 1~ton) is expected~\cite{bib:kamlandzen:nu2014}.

\section{Double Beta Decay in $^{76}$Ge}
\label{sec:ge76}
Germanium became a warhorse of $0\nu\beta\beta$ decay searches once
it was realized that $^{76}$Ge emitter could be embedded in solid state
detectors using a calorimetric approach with High Purity Germanium (HPGe)
diodes. Thanks to their excellent energy resolution (germanium diodes are
still the best gamma spectroscopy detectors to date) in the order
of 0.1-0.2\% FWHM at 2 MeV and to the industrial manufacturing technology,
sizable mass detectors can be built.
Unfortunately, due to the quite low natural abundance of $^{76}$Ge
(7.8\%), isotopically enriched material has to be procured, before
constructing HPGe didoes.
Milestone experiments have been performed by the
Heidelberg-Moscow~\cite{bib:hdm} (HdM) and IGEX~\cite{bib:igex}
collaborations: they used 11~kg and 8~kg of isotopically enriched
(up to 86\%) $^{76}$Ge germanium diodes operated in low activity vacuum
cryostats located in underground laboratories, Laboratori Nazionali del
Gran Sasso (LNGS), in Italy for HdM, and Laboratorio Subterraneo de Canfranc
(LSC), in Spain for IGEX.

The Germanium $Q_{\beta\beta}$ is not very high,
($Q_{\beta\beta} = 2039.061 \pm 0.007~\mbox{keV}$~\cite{bib:ge_qbb})
and lies in a region where contamination from background sources are
possible: apart from the $^{238}$U and $^{232}$Th decay chains that
can contribute to the $0\nu\beta\beta$ region-of-interest (ROI), sizable
contamination can arise from long-lived cosmogenically produced isotopes
($^{68}$Ge and $^{60}$Co in copper and germanium activation) and from
few anthropogenic radioisotopes. Therefore the experiments have to
fight background reduction with careful screening of all the materials
close to the detectors and develop pulse shape discrimination
techniques to further reduce the background contamination.

Part of the HdM collaboration claimed evidence for a peak at $Q_{\beta\beta}$
which corresponds to a half-life central value of
$T_{1/2}^{0\nu} = 1.19 \cdot 10^{25}~\mbox{yr}$~\cite{bib:ge:hdm_value:119}.
The result was later refined with pulse shape analysis (PSA) techniques
giving a half life
$T_{1/2}^{0\nu} = 2.23^{+ 0.44}_{-0.31}\cdot
10^{25}~\mbox{yr}$~\cite{bib:ge:hdm_value:223}.
For a discussion on the two results, see~\cite{bib:2nubb:bernard_review}.

Two larger scale experiment, GERDA\cite{bib:gerda} in Europe
and MAJORANA\cite{bib:majorana} in USA, are
exploiting the germanium diodes technology and beside scrutinizing the
previous claims~\cite{bib:ge:hdm_value:119}-\cite{bib:ge:hdm_value:223}
they will try to push the
experimental sensitivity to the limits.

\subsection{GERDA}
The GERmanium Detector Array (GERDA) experiment operates germanium
diodes made of isotopically modified material, enriched to about
86\% in $^{76}$Ge, without encapsulation in a liquid argon cryogenic
bath. The experiment aims to pursue very low backgrounds thanks
to ultra-pure shielding against environmental radiation.
The germanium detectors are suspended in strings into the cryostat where
64~m$^3$ of LAr are used both as a coolant and shield. The stainless
steel cryostat vessel is covered, from the inside, with copper lining to
reduce gamma radiation from the cryostat walls.
The vessel is surrounded by a large tank filled with high purity
water (590~m$^3$) which further shields the inner volumes from the experimental
hall radiation (absorbing $\gamma$s and moderating neutrons). Moreover it
provides a sensitive medium for the muon system which operates as
a Cerenkov muon veto.

The first phase of the experiment has started on November 9, 2011 using
eight reprocessed coaxial germanium detectors from the HdM and IGEX
experiments together with three natural germanium diodes.
In July 2012, two of the coaxial detectors with natural isotopic abundance
have been replaced by five new enriched Broad Energy Germanium (BEGe)
detectors. The latters are a sub-sample of the thirty new BEGe detectors
recently constructed by Canberra for the Phase II of the experiment.

\begin{figure}[thb]
\begin{center}
\includegraphics[width=10.0cm]{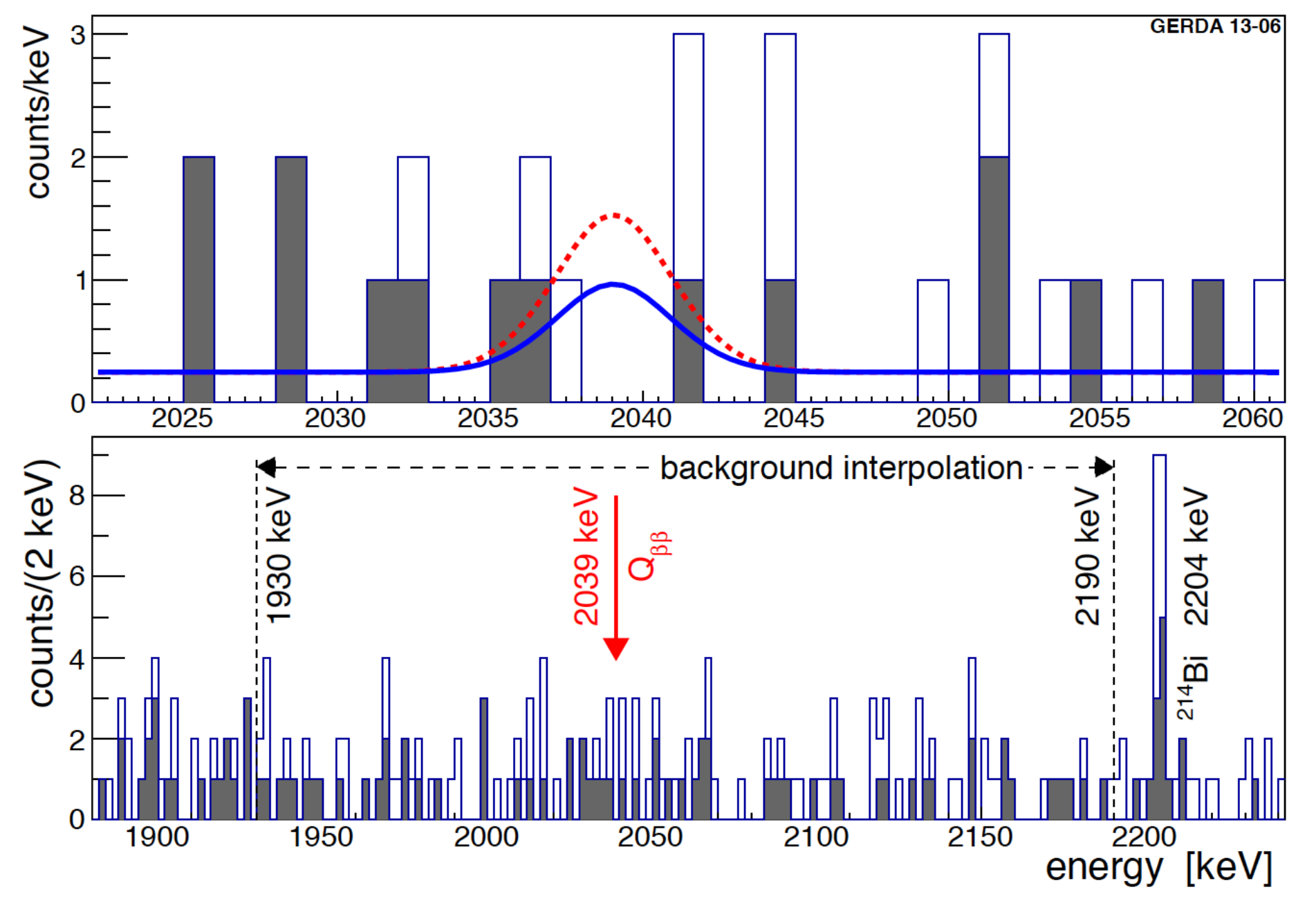}
\end{center}
\caption{Top: Energy spectrum for the sum of all $^{enr}$Ge
detectors in the range 2020-2060 keV, without PSD (empty histogram) and
with PSD (black histogram). The red dashed curve is the expectation based
on the central value of \protect\cite{bib:ge:hdm_value:119}, the blue curve 
is instead based on the 90\% upper limit
derived by \G, $T^{0\nu}_{1/2}$ = 2.1 $\times 10^{25}$ yr.
Bottom: Energy spectrum in the range 1880-2240 keV, the vertical dashed
lines indicate the interval 1930-2190 keV used for the background estimation.
Taken from Ref.~\protect\cite{bib:gerda:0nubb}.}
\label{fig:gerda:spectrum_qbb}
\end{figure}

The first 5.04~kg~yr data, collected before the insertion of the new BEGe
detector, have been used to measure the half-life of the $2\nu\beta\beta$
decay of $^{76}$Ge~\cite{bib:gerda:2nubb}. The extracted half-life is
$$T^{2\nu}_{1/2} = \left ( 1.84^{+0.09}_{-0.08 ~\mbox{fit}} ~~{}^{+0.11}_{-0.06~\mbox{syst}} \right ) \times 10^{21} \mbox{yr.}$$

After having studied the background decomposition of the collected
energy spectra~\cite{bib:gerda:bkg}, keeping the $0\nu\beta\beta$
ROI blinded, Pulse Shape Discrimination techniques have been
developed for coax and BEGe detectors~\cite{bib:gerda:psd}.

The $0\nu\beta\beta$ analysis~\cite{bib:gerda:0nubb}
covers the full data taking, from November 9, 2011 to May 21, 2013,
for a total exposure of 21.60 \expos. Data were grouped into three subsets
with similar characteristics: (1) the golden data set contains the
major part of the data from coaxial detectors, (2) two short run
periods with higher background levels when the BEGe detector were inserted
(silver data set) and (3) the BEGe detectors data set.
Figure.~\ref{fig:gerda:spectrum_qbb} shows the energy spectrum around the
region of interest with and without PSD selection.
No excess of events was observed over a flat background distribution.
Seven events were seen in the range $Q_{\beta\beta}$ $\pm$ 5 keV before
PSD cuts, while 5.1 were expected. 
Only 3 events survived
(classified as Single-Site Events~\cite{bib:gerda:0nubb})
after PSD cuts and no event remained in $Q_{\beta\beta}$ $\pm \sigma_E$.  
To derive the number of signal counts $N^{0\nu}$, a profile likelihood
fit of the spectrum was performed. The fit function was given by the sum
of a constant term for the background and of a Gaussian for the signal events.
The profile likelihood ratio was limited to the region $1/T^{0\nu}_{1/2} > 0$. 
The best fit was obtained for $N^{0\nu}$ = 0, that is no excess of signal
events above background.
The lower limit on the half-life is:
$$T^{0\nu}_{1/2} > 2.1 \times 10^{25} \mbox{yr}~~ @ ~90\% ~~\mbox{C.L. ,}$$
including systematic uncertainties.
This limit corresponds to $N^{0\nu} <$ 3.47 events. 
A Bayesian calculation using   a flat prior distribution for
$1/T^{0\nu}_{1/2}$ from 0 to 10$^{-24}$/yr gives 
$T^{0\nu}_{1/2} > 1.9 \times 10^{25} \mbox{yr}~~ @ ~90\% ~~\mbox{C.I.}$

The data do not show any peak at $Q_{\beta\beta}$ and the result does not 
support the claim described in Ref.~\cite{bib:ge:hdm_value:223}.
The GERDA result is consistent with the negative result from IGEX and
HdM. A combined profile likelihood fit including all these three negative
results gives a 90\% probability limit:
$T^{0\nu}_{1/2} > 3.0 \times 10^{25} \mbox{yr}~~ @ ~90\% ~~\mbox{C.L.}$

A new phase II of the GERDA experiment is planned to start at the end
of 2014, after a shutdown phase prepared to upgrade the experiment
infrastructure and install thirty new BEGe diodes. The detectors, which have
been characterized at the HADES underground
laboratory~\cite{bib:hades:heroica} in Belgium, will increase the experiment
active mass by about 20 kg (of which 3.6~kg, corresponding to 5 BEGe diodes,
were inserted during phase I and their data included in
the published results~\cite{bib:gerda:0nubb}).
Thanks to liquid argon instrumentation and
enhanced pulse shape discrimination power of the new BEGe detectors,
a background reduction from the current BI of about
$2 \cdot10^{-2}~\mbox{cts}/(\mbox{keV kg yr})$ to
$0.1 \cdot10^{-2}~\mbox{cts}/(\mbox{keV kg yr})$ is expected.
With the new configuration, the experiment is supposed to collect
an exposure of about 100~kg~yr and
and improve the $0\nu\beta\beta$
sensitivity to $T_{1/2}^{0\nu} > 1.35 \cdot 10^{26}~\mbox{yr}$.

\subsection{MAJORANA}
While GERDA is preparing the phase II data taking in Europe,
the MAJORANA collaboration in USA
is planning to build a large mass germanium experiment using the
status of the art technology in diode production with a accurate
selection and custom production of radio-pure materials. The proposal
is to mount HPGe diodes inside ultra clean electro-formed copper
vacuum cryostats and place the whole apparatus in a very deep
underground laboratory. Presently, the MAJORANA collaboration is building
a prototype, the MAJORANA demonstrator~\cite{bib:majorana:mjd}
(MJD), to prove the feasibility
of the experiment and measure the background conditions.
The MJD will be constructed
and operated in the
Sanford Underground Research Facility (SURF)
at a depth of 1500~m in South Dakota, USA.
The MJD will use about 40~kg of germanium diodes (with about 30~kg 
of enriched germanium diodes) and, with performances comparable to
those of GERDA, is expected to start data taking in 2014.

Depending on the results of GERDA Phase II and the MAJORANA demonstrator,
a next generation germanium experiment using of the order of 1 ton of
enriched germanium diodes is under discussion. The effort could be
built in stages, starting to merge the GERDA Phase II and MAJORANA
diodes in a common effort environment while constructing new
diodes from enriched material to enlarge the detector active mass.

%

\section{Double Beta Decay in $^{130}$Te}
\label{sec:te128}
Tellurium is another good candidate suitable for $\beta\beta$ decay searches:
due to its high natural abundance (33.8\%) isotopic enrichment is
not needed and it can be used in the form of TeO$_2$ to build bolometric
detectors. Bolometers are calorimeters
operated at milli-kelvin temperatures that can measure the energy released
in the crystal by interacting particles through their temperature rise.
Finally the $Q_{\beta\beta}$ of the decay is relatively high
($Q_{\beta\beta} = 2527~\mbox{keV}$) meaning small background contamination
in the ROI.

\subsection{Cuore}
\label{sec:cuore}
The Cryogenic Underground Detector for Rare Events (CUORE)~\cite{bib:cuore}
is an experiment under construction exploiting a large mass of bolometers:
its design consists of about 1000 natural TeO$_2$ crystals grouped in
19 separated towers. Each crystal is a cube with a side of 5~cm and
a mass of 750~g. The small temperature rise originating from nuclear
decays in the crystals are read using Neutron Transmutation Doped (NTD)
Ge termistors.
The array will be operated at about 10~mK in
a custom He$^3$/He$^4$ dilution refrigerator.
The experiment will be located at LNGS in the same experimental hall
of the GERDA experiment.
The technology has been successfully validated with the
Cuoricino~\cite{bib:cuoricino}
experiment. The first installed CUORE tower, CUORE0 has been configured
as a stand-alone experiment and is currently taking data to study the
background rates and sensitivities expected for CUORE. Recent results
from CUORE0~\cite{bib:cuore:nu2014} show a good energy resolution of
5.2~keV FWHM
at the $^{228}$Tl line (2615~keV) and a background counting rate of
$0.063\pm 0.006$~\ctsper~\cite{bib:cuore:nu2014}.
The full CUORE experiment will start data taking
in 2015 with an expected sensitivity on $0\nu\beta\beta$
of $2.1\times 10^{26}~\mbox{yr}$.

\section{Conclusions}
Neutrinoless double beta decay is an exciting physics topic and
double beta decay searches keep on playing a unique role in neutrino
physics: probing the lepton number conservation, they can shed light
on the Dirac/Majorana nature of neutrinos and indirectly measure the absolute
neutrino mass scale with high sensitivity.
Several large mass, high sensitivity, experiments will be running in the next
few years and they will provide important results on $0\nu\beta\beta$.
In case of
a positive signal, an observation with several isotopes is needed for
convincing evidence. The results would imply that neutrino follow
an inverted hierarchy~\footnote{or degenerate, or normal if there are sterile
neutrinos. The argument is far from simple and many theoretical scenarios
are still open.}
mass scheme and allow to directly measure the neutrino
mass scale.
Even a missing observation of $0\nu\beta\beta$ on all the isotopes under
investigation would play an important role and the results would have
to be combined with those coming from future neutrino oscillation
experiments (reactors and long baseline).

Concerning the direct hierarchy mass scheme, at present none of the
experiments seems to have any reasonable chance of going below
the inverted hierarchy scheme. Therefore, new strategies and revolutionary
techniques would have to be
developed to push further the experimental sensitivities.







\end{document}

%% file: econfmacros.tex
\def\beq{\begin{equation}}
\def\eeq#1{\label{#1}\end{equation}}
\def\eeqn{\end{equation}}

\def\beqa{\begin{eqnarray}}
\def\eeqa#1{\label{#1}\end{eqnarray}}
\def\eeqan{\end{eqnarray}}

\let\bar=\overbar

\def\Dslash{\not{\hbox{\kern-4pt $D$}}}
\def\dslash{\not{\hbox{\kern-2pt $\del$}}}

\def\msb{{\bar{\ssstyle M \kern -1pt S}}}




%% file: ag_dbexp_fpcp2014.bbl
\begin{thebibliography}{99}

\bibitem{bib:2nubb_pdg}
J.\ Beringer {\it et al.}, Particle Data Group,
{\it Review of Particle Physics, Phys. Rev.} {\bf D\/86} (2012).
%
\bibitem{bib:2nubb:rev_bara}
A.\ S.\ Barabash, {\it Phys. Rev.} {\bf C\/81}, 035501 (2010),
[\arXiv{1003.1005}]

\bibitem{bib:0nubb:rode_standard_inter}
W.\ Rodejohann 2011 {\it Int. J. Mod. Phys.} {\bf E\/20} 1833 (2011),
[\arXiv{1106.1334}]

\bibitem{bib:2nubb:bernard_review}
B.\ Schwingenheuer, {\it Ann. Phys.} {\bf 525} 269 (2013),
[\arXiv{1210.7432}]

\bibitem{bib:2nubb:rodejohann_review}
P.\ S.\ Bhupal Dev, et al {\it Phys. Rev.} {\bf D 88} 091301(R) (2013),
[\arXiv{1305.0056}]

\bibitem{bib:EXO_jinst}
M.\ Auger et al., {\it JINST} {\bf 7} P05010 (2012),
[\arXiv{1202.2192}]


\bibitem{bib:EXO_0nubb}
EXO Collab. (M.\ Auger {\it at al}.),
{\it Phys. Rev. Lett.} {\bf 109} 032505 (2012),
[\arXiv{1205.5608}]

\bibitem{bib:EXO_0nubb:nature}
EXO Collab. (J.\ B.\ Albert {\it at al}.),
{\it Nature} {\bf 510} 229 (2014),
[\arXiv{1402.6956}]

\bibitem{bib:EXO_2nubb_second}
EXO Collab. (J.\ B.\ Albert {\it et al}.),
{\it Phys. Rev.} {\bf C 89} 015502 (2014),
[\arXiv{1306.6106}].

\bibitem{bib:nEXO}
G.\ Gratta, D.\ Sinclair, {\it Adv.High Energy Phys.} 2013 545431 (2013).

\bibitem{bib:kamland_jinst}
KamLAND Collab. (S.\ Abe {\it et al}.),
{\it Phys. Rev.} {\bf C 81} (2012) 025807,
[\arXiv{0907.0066}].

\bibitem{bib:kamlandzen:2nubb}
KamLAND-Zen Collab. (A.\ Gando {\it et al}.),
{\it Phys. Rev.} {\bf C 85} 045504 (2012),
[\arXiv{1201.4664}].

\bibitem{bib:kamlandzen:0nubb}
KamLAND-Zen Collab. (A.\ Gando {\it et al}.),
{\it Phys. Rev. Lett.} {\bf 110} 062502 (2013),
[\arXiv{1211.3863}].

\bibitem{bib:kamlandzen:nu2014}
A.\ Shimizu, {\it Results from KamLAND-Zen},
presentation at the XXVI Conference on Neutrino Physics and
Astrophysics, June 2-7, 2014, Boston , USA.


\bibitem{bib:hdm}
HdM Collab. (H.\ V.\ Klapdor-Kleingrothaus {\it at al}.),
{\it Eur. Phys. J.} {\bf A 12} 147 (2001),
[\hepph{0103062}].

\bibitem{bib:igex}
IGEX Collab. (C.\ E.\ Aalseth {\it et al}.),
{\it Phys. Rev.} {\bf D 65} 092007 (2002),
[\hepex{0202026}].

\bibitem{bib:ge_qbb}
B.\ J.\ Mount, M.\ Redshaw and E.\ G.\ Mayers,
{\it Phys. Rev.} {\bf C 81} 032501 (2010).

\bibitem{bib:ge:hdm_value:119}
H.\ V.\ Klapdor-Kleingrothaus, {\it et al}.
{\it Phys. Lett.} {\bf B 586} 198 (2004),
[\hepph{0404088}].

\bibitem{bib:ge:hdm_value:223}
 H.\ V.\ Klapdor-Kleingrothaus and I.\ V.\ Krivosheina~I.~V.
{\it Mod. Phys. Lett.} {\bf A 21} 1547 (2006).


\bibitem{bib:gerda}
GERDA Collab. (K.\ H.\ Ackermann {\it et al}.)
{\it Eur. Phys. J.} {\bf C 73} 2330 (2013),
[\arXiv{1212.4067}].

\bibitem{bib:majorana}
MAJORANA Collab. (R.\ Gaitskell {\it et al}.),
{\emph White paper on the Majorana zero-neutrino double-beta decay
experiment} [\nuclex{0311013}]

\bibitem{bib:gerda:2nubb}
GERDA Collab. (M.\ Agostini {\it et al}.),
{\it J. Phys. G: Nucl. Part. Phys.} {\bf 40} 035110 (2013),
[\arXiv{1212.3210}].

\bibitem{bib:gerda:bkg}
GERDA Collab. (M.\ Agostini {\it et al}.),
{\it Eur. Phys. J.} {\bf C 74} 2764 (2014),
[\arXiv{1306.5084}]

\bibitem{bib:gerda:psd}
GERDA Collab. (M.\ Agostini {\it et al}.),
{\it Eur. Phys. J.} {\bf C 73} 2583 (2013),
[\arXiv{1307.2610}]

\bibitem{bib:gerda:0nubb}
GERDA Collab. (M.\ Agostini {\it et al}.),
{\it Phys. Rev. Lett.} {\bf 111} 122503 (2013),
[\arXiv{1307.4720}]

\bibitem{bib:hades:heroica}
E.\ Andreotti {\it et al}., {\it JINST} {\bf 8} P06012 (2013),
[\arXiv{1302.4277}]

\bibitem{bib:majorana:mjd}
MAJORANA Collab. (R.\ D.\ Martin {\it et al}.), 
{\it Status of the MAJORANA demonstrator} [\arXiv{1311.3310}]


\bibitem{bib:cuore}
CUORE Collab. (C.\ Arnaboldi {\it et al}.)
{\it Nucl. Instr. and Meth.} {\bf A 518} 775 (2004)
[\hepex{0212053}] \\
CUORE Collab. (D.\ R.\ Artusa {\it et al}.),
{\emph Searching for neutrinoless double-beta decay of $^{13}$Te with
CUORE}, [\arXiv{1402.6072]}

\bibitem{bib:cuoricino}
C.\ Arnaboldi {\it et al}.
{\it Phys. Rev.} {\bf C 78} 035502 (2008)

\bibitem{bib:cuore:nu2014}
O.\ Cremonesi, {\it CUORE0 results and prospects for the CUORE experiment},
presentation at the XXVI Conference on Neutrino Physics and
Astrophysics, June 2-7, 2014, Boston , USA.

\end{thebibliography}
